\documentstyle[aps]{revtex}
\draft
\input epsf
\begin{document}
\twocolumn[\hsize\textwidth\columnwidth\hsize\csname@twocolumnfalse%
\endcsname
\title{Absence of a Finite-Temperature Melting Transition in the
Classical
       Two-Dimensional One-Component Plasma}
\author{M.\ A.\ Moore$^1$ and A.\ P\'erez--Garrido$^2$}

\address{$^{1\,}$ Theoretical Physics Group,
    Department of Physics and Astronomy,
   The University of Manchester, M13 9PL, UK}
\address{$^{2\,}$ Departamento de F\'{\i}sica, Universidad de Murcia,
Murcia 30.071, Spain}
\maketitle
\begin{abstract}
{Vortices in thin-film superconductors are often modelled as a system 
of
particles interacting via a repulsive logarithmic potential.  
Arguments are
presented to show that the hypothetical (Abrikosov) crystalline state 
for such
particles is unstable at any finite temperature against proliferation 
of
screened  disclinations.  The correlation length of crystalline order 
is
predicted to grow as $\sqrt{1/T}$ as the temperature $T$ is reduced 
to zero, in
excellent agreement with our simulations of this two-dimensional 
system. 
}
\end{abstract}
]
\narrowtext

It has been commonly assumed for many years now that for the 
physically
important case of particles moving in two dimensions interacting with
each
other via a repulsive logarithmic potential (a situation sometimes
called the
two-dimensional one-component plasma problem) one would have the 
usual
phases
expected on the KTHNY scenario. This scenario describes 
two-dimensional 
melting
as a defect--mediated phenomenon (Halperin, Nelson \cite{HN78} and 
Young
\cite{Y79}) and is based on ideas of Kosterlitz and Thouless
\cite{KT73}.  It
is supposed that the crystalline phase -- a triangular lattice -- 
melts
at a
continuous transition into an hexatic liquid  due to the 
proliferation
of
dislocations. The hexatic liquid becomes an ordinary liquid at
temperatures
which permit the creation of disclinations. This scenario is
well-established
for particles with short-range interactions \cite{CL}, but we will
argue  that 
particles interacting with a logarithmic potential behave completely 
differently. For them  we can show that the crystalline state is
unstable at
any temperature against the proliferation of (screened) disclinations
and as a
consequence the system stays in the liquid state down to arbitrarily 
low
temperatures. The ground state of the system is of course 
crystalline;
the
correlation length of short-range crystalline order is predicted to
grow  as
$\sqrt{1/T}$ as the temperature $T$ approaches zero. Our numerical
simulations
reported here confirm this behavior.

The one-component plasma problem is of considerable  physical
significance as
it relates to the thermodynamics of vortices --``the particles" -- in
thin 
film
superconductors. For thin enough films the screening length in the
intervortex
potential may be greater than the transverse dimensions of the film,
which
makes the logarithmic potential   an accurate approximation for the
potential. 
Most papers on thin film superconductors assume 
the vortices have a freezing transition at low enough temperatures 
(for
a
review see Ref. \cite{Blatter}), although clear experimental evidence
for
this is lacking. For a contrary view, however, see \cite{DM97} and
references
cited therein.

For particles interacting with a repulsive potential a device  is 
needed
to
stop them escaping to infinity. In numerical studies of 
two-dimensional 
melting
the most commonly used device is  periodic boundary conditions.
Unfortunately
the use of this boundary condition  with either short-range 
interactions
or
with the logarithmic interaction \cite{perram} produces  an 
apparently
first
order transition between the crystal and  liquid states rather than 
the
KTHNY
scenario. (For  a review of early work on short-range interactions 
see
\cite{St88}; for some more recent work see  Refs.  \cite{BA96} or
\cite{Jaster}).  This is probably a finite size effect: studies on
systems 
with
over 60,000 particles indicate that the van der Waals loops 
associated
with 
the
apparent first-order transition shrink in these very large systems 
\cite{BA96},
\cite{Jaster}. We ourselves have found  that placing the particles on
the
surface of a sphere is very effective for short-range interactions 
\cite{GM98}:
no van der Waals loops occur with this topology and the results 
obtained
even 
with   modest numbers of particles  are in excellent agreement with
expectations based on KTHNY theory. As a consequence the numerical 
work
which
we are reporting in this paper has been carried out for  the
two-dimensional
system  represented by  the surface of a sphere. 

The ground state configuration of the particles on the sphere
has to contain at least 12 disclinations
(5-fold rings) by Euler's theorem. We have made extensive studies of
these
ground states and discovered that for larger systems the 
disclinations
are
screened by lines of dislocations \cite{DM97}, \cite{apg97}.  These
defects
within the crystalline state seem to overcome the problem of the
spurious 
first order transition induced by finite size effects when periodic
boundary
conditions are employed and so enable one to get results closer to 
those
obtaining in the thermodynamic limit. It is  noteworthy that an early
simulation of the  one-component plasma on the
surface of a sphere \cite{Caillol}  did not find  a finite 
temperature 
phase 
transition either, in agreement with our results. 

The Hamiltonian for  particles moving on the surface of the sphere
interacting via a logarithmic potential is

\begin{equation}
H=-J\sum_{i < j}ln(|{\bf r}_i-{\bf r}_j|/R),
\label{hamiltonian}
\end{equation} 
where ${\bf r}_i$ is the position of the $i$th particle on the 
surface
of the
sphere, $R$ is the radius of the sphere and $J$ is a measure of the
magnitude
of the repulsive forces between the particles. The key feature which
distinguishes the logarithmic potential from other potentials is that
all the
stationary states of $H$ have zero dipole moment \cite{bbp}. This was
proved
 by noting that the force on the $i$th particle due to all the others
must be directed radially for any equilibrium configuration since
otherwise 
the
particle would move along the sphere, so

\begin{equation}
\sum_{j \ne i}\frac{{\bf r}_i-{\bf r}_j}{|{\bf r}_i-{\bf r}_j|^2}=
f_i{\bf 
r}_i.
\label{stationary}
\end{equation}
By multiplying both sides by ${\bf r}_{i}$ one can show that 
$f_{i}=(N-1)/2R^2$
where $N$ is the total number of  particles. By summing Eq.
(\ref{stationary})
over all $i$ and using the fact that ${\bf r}_{i}-{\bf r}_j$ is
antisymmetric
in $i$ and $j$, it follows that the dipole moment, $\sum_{i} {\bf 
r}_i$,
is
zero. No other potential has this feature and it has important
consequences.

\begin{figure}
\epsfxsize=.8\hsize
\begin{center}
\leavevmode
\epsfbox{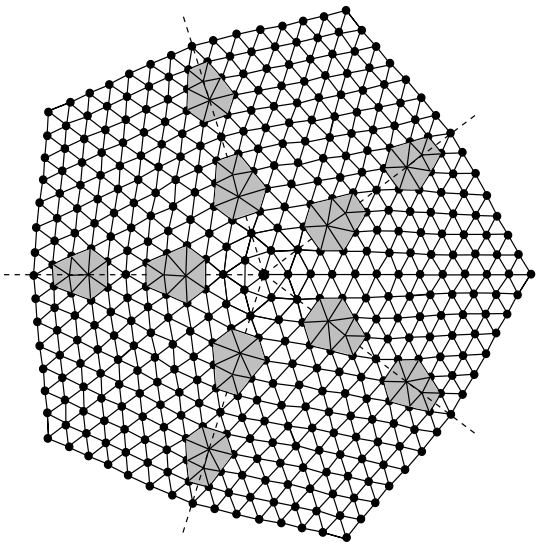}
\end{center}
\caption{Dislocation series screening a five-fold disclination.}
\label{dislo}
\end{figure}

Our basic contention is that thermally excited screened disclinations
will
destroy crystalline order for particles interacting with each other 
via
a
logarithmic potential. The energy cost of an unscreened disclination 
is  
$O(N)$
\cite{CL} and such a disclination will not be thermally created in a
crystalline state. However, disclinations can be ``screened" by a 
cloud
of
dislocations and it turns out that the energy cost of such screened
disclinations can be  much smaller, of $O(ln(N))$ for non-logarithmic
potentials and $O(1/N)$ for the logarithmic potential. The phenomenon 
of
screening of a disclination by dislocations is well-known \cite{CL},
\cite{DM97}, and is illustrated in Fig. 1. The figure shows a 
five-fold
coordinated ring -- a disclination  -- screened by five lines of 
dislocations
where the  dislocations are spaced a distance $cl_0$ apart; $l_0$ is 
the 
lattice spacing. The strain field of the central disclination can be 
largely
cancelled by that arising from the lines  of dislocations. The 
resulting strain
field from the dislocations along a line 
may
be approximated at large distances by  that which arises from a 
positive
and
negative disclination at each end of the line with a ``disclination" 
charge 
of
size $q_s=1/c$ (where $q_s=+1$ for a fivefold disclination). If we
consider a
line of dislocations with $c=5$ and five lines as in Fig. 1 then the
strain
field of the central disclination is exactly screened away. As shown 
in
Ref.
\cite{DM97} the contribution of the disclinations at the other end of
the 
lines
can be  made arbitrarily small by allowing the spacing of the
dislocations
to increase with distance $r$ from the disclination as
$c(r)=5+S/l_0g(r/S)$
with the condition that $g(0)=0$; $S$ is the size or scale of the
screened
disclination. Then  the residual  charge associated with the screened
disclination can be made as small as $O(l_0/S)$ but, as we shall  
show,
must 
be
as small as $O((l_0/S)^2)$ for the special case of the logarithmic
potential. 

First let us review some features of two-dimensional continuum
elasticity
theory \cite{CL}. Small strains $u_{ij}({\bf r})$ are related to the
stress
field by Hooke's law,
$\sigma_{ij}=B\delta_{ij}u_{kk}+2\mu(u_{ij}-\delta_{ij}u_{kk}/2)$, 
where
$B$ 
is
the bulk modulus and $\mu$ is the shear modulus. In the presence of 
topological
defects it is convenient to introduce the Airy stress function, 
$\chi$, 
defined
by $\sigma_{ij}=\epsilon_{ik}\epsilon_{jl}\partial_k\partial_l\chi$. 
A 
fivefold
disclination  is defined by a change in bond angle $2\pi/6$ when a
path encircles the defect. Dislocations are defined by their Burgers
vector
density field ${\bf b}({\bf r})$ which for the dislocations in Fig. 1
points
perpendicular to the line upon which they lie. The stress field is
related to
the densities of disclination $s({\bf r})$ and dislocations via

\begin{equation}
\frac{1}{Y_2}\nabla^4\chi=s({\bf r})-\epsilon_{ik}\nabla_kb_i({\bf
r})\equiv
\tilde{s}({\bf r}),
\label{Airy}
\end{equation}
where $Y_2 =4B\mu/(B+\mu)$. For a single disclination at the origin 
as
in Fig.
1, $s({\bf r})=(2\pi/6)\delta({\bf r})$. $\tilde{s}({\bf r})$ can be
regarded as a total disclination density made up of a ``free"
disclination
density $s({\bf r})$ and a ``polarization" contribution
$-\epsilon_{ik}\nabla_kb_i$ from dislocations.  The energy of the
screened
disclination expressed  in Fourier space is

\begin{equation}
E=\frac{1}{2}Y_2\int\frac{d^2q}{(2\pi)^2}\frac{1}{q^4}\tilde{s}({\bf
q})\tilde{s}(-{\bf q}).
\label{energy}
\end{equation}
The  expectation \cite{DM97} for 
non-logarithmic interaction potentials is that the screening can at 
best
make the Fourier transform of $\tilde{s}$, $\tilde{s}({\bf
q})=ql_0f(qS)$ when
 the amount of ``disclination charge" within a region of radius $S$
around
the  disclination is of $O(l_0/S)$. 
Substituting this form for $\tilde{s}({\bf q})$ into
Eq. (\ref{energy}), one finds that in a system of $N$ particles
the energy of the screened disclination would be of $O(lnN)$ -- which
explains
why screened disclinations would be
unlikely to modify the KTHNY scenario for non-logarithmic 
interactions. 

The  change in the particle density  due to the
presence of the topological defects is, when Fourier transformed, 
given
by

\begin{equation}
\Delta\rho({\bf q})=-nu_{ii}({\bf q})=\frac{n}{2B}q^2\chi({\bf q}),
\label{density}
\end{equation}
where $n$ is the number density of the particles.  A feature of the 
logarithmic
potential is that for it the bulk modulus $B$ is not  constant
\cite{Brandt77} 
but diverges at small wavevector: $B(q)=2\pi Jn^2/q^2$. The shear
modulus is
well-behaved: $\mu=Jn/8$. Equations (\ref{energy}) and 
(\ref{density})
can
still be used if the replacements $Y_2\rightarrow 4\mu$ and
$B\rightarrow 
B(q)$
are employed.  At small wave-vector it follows that for the 
logarithmic
potential

\begin{equation}
\Delta\rho({\bf q})=\frac{1}{8\pi}\tilde{s}({\bf q}).
\label{rhos}
\end{equation}  
Only for the logarithmic case  does  a finite small $q$ limit exist 
for
the
density change $\Delta\rho$ associated with a screened disclination.  

We now exploit the fact that all stationary states of $H$ have 
vanishing 
dipole
moment to show that for the case of logarithmic interactions between 
the
particles the screening of the disclination is more efficient than 
for
non-logarithmic interactions.  The Fourier transform of the
particle density is defined by

\begin{equation}
\rho({\bf q})=\frac{1}{N}\sum_ie^{i{\bf q}.{\bf r}_i}.
\label{defndensity}
\end{equation}
(Formally, as our system is the surface of a sphere rather than  a 
plane we
should use spherical harmonics rather than plane waves, as was done 
in
Ref.\cite{DM97}, but  the distinction is unimportant for our 
argument).  
Consider now a state which differs from the groundstate by  the 
presence of 
a screened disclination of size $S$. The density difference 
$\Delta\rho({\bf
q})$ of  the two states must differ as $q\rightarrow0$ by terms of 
$O(q^2)$ ;
(if one of the states had had a dipole moment then one can see from  
expanding
the  exponential in Eq. (\ref{defndensity}) that $\Delta\rho$ would 
have been
of $O(q)$). Eq. (\ref{rhos}) implies that $\tilde{s}({\bf 
q})=q^2l_{0}^2f(qS)$.
By Fourier transforming $\tilde{s}({\bf q})$ one can then show that 
the
``disclination charge" within a distance S of the center of the 
disclination is
of $O((l_0/S)^2)$.

Substituting this form for $\tilde{s}$ into Eq. (\ref{energy}) one 
finds
that
the energy of the screened disclination is of order $J(l_0/S)^2$. By 
increasing
the scale S it can be made arbitrarily small. At a temperature $T$ a
region of
linear extent $\xi$, where $J(l_0/\xi)^2=T$, will be unlikely to 
contain
a
disclination and so $\xi$ will be a measure  of the short-range
crystalline
order present in the system at temperature $T$. $\xi$ diverges as
$\sqrt{1/T}$
as $T\rightarrow0$. This means that by investigating numerically  the
structure factor  one can find from the widths of its peaks the
 correlation length $\xi$ and its temperature dependence will tell us
whether
the arguments above are valid.

We studied using molecular dynamics, specifically a velocity Verlet 
algorithm
\cite{FS96}, a system of $N$ particles confined to move on the 
surface of a
sphere and interacting with the logarithmic potential. Reduced units 
were used,
i.e. $m=k_{\rm B}=R=J=1$, where $m$ is the mass of the particle and 
$k_{\rm B}$
is  the Boltzmann constant. The acceleration ${\bf a}_i$ of the $i$th 
particle
equals ${\bf f}_i/m$,  where ${\bf f}_i$ is the force produced by the 
other
particles on the $i$th particle. After a small time interval $\delta 
t$ the
position of the particle will be  ${\bf x}_i={\bf r}_i(t)+{\bf 
v}_i(t)\delta
t+\frac{1}{2}{\bf a}_i(t)\delta t^2$, where ${\bf v}_i(t)$ is the 
velocity of
the particle. In general ${\bf x}_i$ will not lie on the surface of 
the sphere.
The $i$th particle is brought back to the surface by acting on it 
with a 
fictitious force $-2\lambda_i{\bf r}_i(t)$ where
\begin{equation}
\lambda_i=\frac{{\bf r}_i(t)\cdot{\bf x}_i-\sqrt{\left[{\bf
r}_i\left(t\right)\cdot
{\bf x}_i\right]^2
-R^2\left[\left|{\bf x}_i\right|^2-R^2\right]}}
{R^2\,\delta t^2}.
\end{equation}
Then, the velocity Verlet algorithm updates particle positions using
the equation:
\begin{equation}
 {\bf r}_i(t+\delta t) = {\bf x}_i-\lambda_i{\bf r}_i(t)\delta t^2.
\end{equation}

We chose $\delta t=0.005(mR^2/J)^{1/2}$. The velocities of the 
particles were
chosen from a Boltzmann distribution appropriate to the temperature 
$T$ and
were re-selected at equally spaced time intervals \cite{AS83}.
The system was equilibrated at high temperatures, then the 
temperature was
slowly reduced.
For each temperature 
 we determined the
structure factor which is related to the Fourier transform of the 
pair
correlation
function $h(r)$ by
\cite{HL79}:
\begin{equation}
S(q)=1+2\pi\rho R^2 \int_0^\pi h(R\theta )\sin\theta J_0(qR\theta
)d\theta ,
\end{equation} 
where $J_0$ is the Bessel function of zeroth order. This adaption of 
the
conventional relation to particles moving on the surface of the 
sphere is valid
provided $q$ is of $O(1)$ and not of $O(1/R)$. Peaks in the structure 
factor
grow as the temperature is reduced at  wavevectors $q$ corresponding 
to the
reciprocal lattice vectors $|{\bf G}|$ of the triangular lattice 
expected for
the groundstate. The correlation length, $\xi$, is the inverse of the 
width of
the  first peak of the structure factor. To determine it, we fitted 
the first
peak to a Lorentzian curve.
\begin{figure}
\epsfxsize=.9\hsize
\begin{center}
\leavevmode
\epsfbox[12 50 576 438]{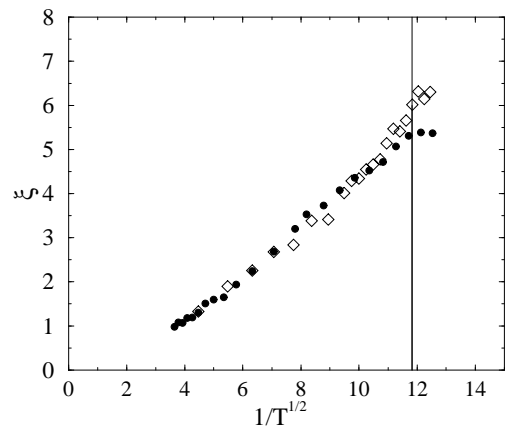}
\end{center}
\caption{Correlation length as a function of $T^{-1/2}$ for 1442
particles (solid circles)
and for 2252 particles (empty diamonds) for the logarithmic 
potential.  }
\label{xi}
\end{figure}

We studied systems of 1442 and 2252 particles.  The simulation time 
for each
temperature was 100,000$\delta t$. In Fig.\ \ref{xi} $\xi$ is plotted 
against
$\sqrt{1/T}$. The vertical line is drawn  where other authors found a
first-order melting transition using periodic boundary conditions
\cite{perram}.  The predicted behavior $\xi\propto \sqrt{1/T}$ is 
clearly seen
in Fig.\ \ref{xi}. When the temperature was reduced to  $T=0.01$ the
correlation length for the system of 1442 particles reached  the 
system size
and stopped growing as the temperature was reduced further. However, 
the
simulation with 2252 particles indicates  that this levelling off is 
just a
finite size effect. True equilibration in numerical studies of 
two-dimensional
melting phenomena is always problematic \cite{St88} and  may be the 
cause of
the scatter in Fig.\ \ref{xi}.

\begin{figure}
\epsfxsize=.9\hsize
\begin{center}
\leavevmode
\epsfbox[12 45 576 438]{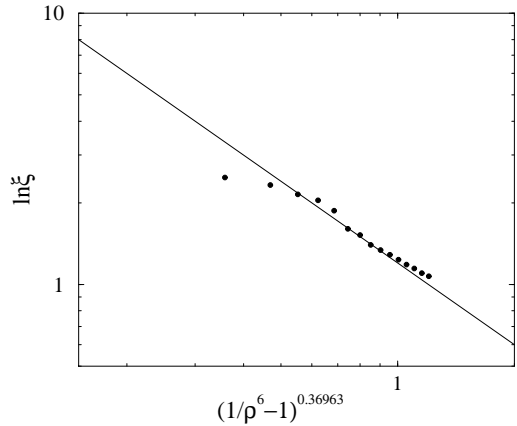}
\end{center}
\caption{Log-log plot of $ln\xi$  versus $(1/\rho^6-1)^{0.36963}$ for
5882 particles interacting with a $1/r^{12}$
repulsive potential. The slope of the straight
line is $-1$ according to KTHNY expectations. }
\label{xi2}
\end{figure}

The apparent absence of a finite temperature phase transition for 
particles
interacting with a logarithmic potential cannot be attributed to the 
fact that
we have done the simulation for the two-dimensional geometry 
represented by the
surface of a sphere. We can demonstrate this by simulating particles
interacting with a $1/r^{12}$ potential also moving on the surface of 
a sphere.
A finite temperature melting transition of  KTHNY character is seen.
We had already found indications of such a  melting transition 
\cite{GM98}
but the  numbers of particles used in that reference  were rather 
small.
Using the Verlet algorithm described above  we are able to simulate
much larger
systems eg. 5882 particles. The calculations for the short-range 
potentials run
faster than with the logarithmic potential as it is possible to use 
look-up
tables of nearest neighbors.
In the KTHNY picture, the correlation length has the following 
density 
dependence along an isotherm for a $1/r^{12}$ potential \cite{BG82}:
\begin{equation}
\xi(\rho)\propto \exp\left( \frac{b}{\left((\rho_c/\rho)^6-1
\right)^\nu}\right),
\label{iso}
\end{equation}
where $\rho$ is the density and $\nu=0.36963\ldots$. In Fig.\ 
\ref{xi2}, 
ln$\xi$  is plotted versus $(1/\rho^6-1)^{0.36963}$. We have assumed  
that
$\rho_c=1$, a value   obtained by other authors \cite{BG82,BA96} 
working at the
temperature which we used,  $T=1$. 
The slope of the straight
line is $-1$ which corresponds well with KTHNY predictions. Note 
again that
 finite size effects cut off the growth of $\xi$ when it is of 
$O(R)$. 
Thus  simulations on the sphere do produce for a non-logarithmic 
potential
 the expected  crystalline phase.

In summary, we have shown that thermal excitation of screened 
disclinations
removes at non-zero temperature  the crystalline phase of the vortex 
system. 
Numerical simulations have confirmed our  prediction that as the 
temperature is
lowered the correlation length of short-range crystalline order 
should grow as
$\xi\propto\sqrt{1/T}$.

APG  would like to acknowledge a grant and financial support from the
Direcci\'on
General de Investiga\-ci\'on Cien\-t\'{\i}\-fica y T\'ec\-nica, 
project
number
PB 96/1118 and EPSRC under grant GR/K53208. We have had useful
discussions 
with
H. Bokil.

\end{document}